\documentclass[superscriptaddress,twocolumn,showpacs,amsmath,amssymb,floats,pre]{revtex4}

\usepackage{graphicx}

\usepackage{dcolumn}

\usepackage{bm}

\usepackage{revsymb}

\usepackage[usenames]{color}

\usepackage{subfigure}

\usepackage{units}

\begin{document}

\newcommand{\brm}[1]{\bm{{\rm #1}}}
\newcommand{\sgn}{{\rm sgn}}
\newcommand{\Tr}{{\rm Tr}}
\newcommand{\tens}[1]{\underline{\underline{#1}}}
\newcommand{\mm}{\overset{\leftrightarrow}{m}}
\newcommand{\xv}{\bm{{\rm x}}}
\newcommand{\Rv}{\bm{{\rm R}}}
\newcommand{\uv}{\bm{{\rm u}}}
\newcommand{\vv}{\bm{{\rm v}}}
\newcommand{\nv}{\bm{{\rm n}}}
\newcommand{\Nv}{\bm{{\rm N}}}
\newcommand{\ev}{\bm{{\rm e}}}
\newcommand{\dv}{\bm{{\rm d}}}
\newcommand{\bv}{\bm{{\rm b}}}
\newcommand{\lv}{{\bm{l}}}
\newcommand{\rv}{\bm{{\rm r}}}
\newcommand{\id}{\tens{\mathbb I}}
\newcommand{\bhv}{\hat{\bv}}
\newcommand{\bh}{\hat{b}}
\def\ten#1{\underline{\underline{{#1}}}}
\newcommand{\Ft}{{\tilde F}}
\newcommand{\Ftv}{\tilde{\mathbf{F}}}
\newcommand{\sigmat}{{\tilde \sigma}}
\newcommand{\sigmab}{{\overline \sigma}}
\newcommand{\ellv}{\mathbf{\ell}}
\newcommand{\qv}{\bm{{\rm q}}}
\newcommand{\pv}{\bm{{\rm p}}}
\newcommand{\tD}{\underline{D}}
\newcommand{\Tchange}[1]{{\color{red}{#1}}}
\newcommand{\Fa}{{\cal F}}
\newcommand{\Uv}{{\roarrow U}}
\newcommand{\Bv}{{\roarrow B}}
\newcommand{\Dt}{{\tensor {\cal D}}}
\newcommand{\cv}{{\mathbf c}}
\newcommand{\pt}{\tilde{p}}
\newcommand{\as}[1]{{#1}}
\newcommand{\fbz}{1$^{st}$ BZ}
\newcommand{\BL}[1]{{#1}}
\newcommand{\re}{{{\rm Re}}}
\newcommand{\im}{{{\rm Im}}}

\title{Emergent tilt order in Dirac polymer liquids}

\author{Anton Souslov}
\affiliation{School of Physics,
Georgia Institute of Technology, Atlanta, GA, 30332, USA }

\author{Benjamin Loewe}
\affiliation{School of Physics,
Georgia Institute of Technology, Atlanta, GA, 30332, USA }

\author{Paul M.~Goldbart}
\affiliation{School of Physics,
Georgia Institute of Technology, Atlanta, GA, 30332, USA }

\date{\today}

\begin{abstract}
\noindent 
We study a liquid of zigzagging two-dimensional directed polymers with bending 
rigidity, i.e., polymers whose conformations follow checkerboard paths. In the continuum limit the statistics of such polymers obey the Dirac equation for particles of imaginary mass. 
We exploit this observation to investigate a liquid of these polymers via a quantum many-fermion analogy. A self-consistent approximation predicts a phase of tilted order, in which the polymers may develop a preference to zig rather than zag. We compute the phase diagram and key response functions for the polymer liquid, and comment on the role played by fluctuations.
\end{abstract}


\maketitle

Directed line liquids consist of quasi-one-dimensional objects that are preferentially oriented along a common direction about which they undergo thermal fluctuations~\cite{DeGennes1968,Fisher1984,LeDoussal1991,Kamien1992}.
Realizations include systems as diverse as 
polymer liquids under uniaxial tension~\cite{Rocklin2012,Souslov2013,Rocklin2013}, 
two-dimensional lamellar smectics~\cite{Golubovic1989}, 
step edges on crystal surfaces~\cite{Bartelt1990}, 
interfaces in the KPZ universality class~\cite{Kulkarni2013}, 
and vortex lines in planar type-II superconductors~\cite{Polkovnikov2005}.
Quantum many-body physics provides powerful tools for analyzing the thermal equilibrium properties of classical systems of strongly interacting directed line liquids, by means of the well-known mapping between the configurations of classical directed lines in $D$-dimensional space and the world-lines of nonrelativistic quantum particles moving in $D-1$ spatial dimensions~\cite{DeGennes1968,Rocklin2012,Souslov2013,Rocklin2013}. 
For example, in Refs.~\cite{DeGennes1968,Rocklin2012,Rocklin2013} this mapping has been used to relate the structure of non-intersecting, but otherwise non-interacting, two-dimensional directed polymer liquids to the properties of the non-interacting one-dimensional Fermi gas.
As discussed in Ref.~\cite{Souslov2013}, the properties of three-dimensional directed polymer liquids follow from a mapping onto a two-dimensional system of fermions interacting via a Chern-Simons potential.

The two-dimensional directed polymers that we consider in the present article have an energetic
preference to be straight, i.e., an energy cost to be \emph{bent}. 
To date, by contrast, attention has primarily been focused on directed polymers that have an energetic
preference to be \emph{short}, i.e., an external tension controls the mean length.
Unlike the Kratky-Porod~\cite{Kratky1949} model, which features a curvature-based bending energy, the polymers in the current model 
bend only by a fixed angle. 
This model of zigzagging directed lines may also capture the effect of crystalline anisotropy in step edges on crystal surfaces.
As we shall see, the appropriate quantum analog of the zigzagging directed line liquid consists of \emph{relativistic} quantum particles,
which, accordingly, are governed by the Dirac equation.
This analogy, combined with a self-consistent field approximation, enables us to obtain information about local polymer density and alignment in the form of
the mean values and correlations of these quantities. 
\textit{Inter alia}, we shall also see that the interplay of bending rigidity and repulsive interactions 
leads to polymer alignment.
This kind of interplay has long been known to promote striking collective phenomena, such as
nematic~\cite{Pincus1978} or smectic~\cite{Golubovic1989} ordering.

\begin{figure}
     \includegraphics{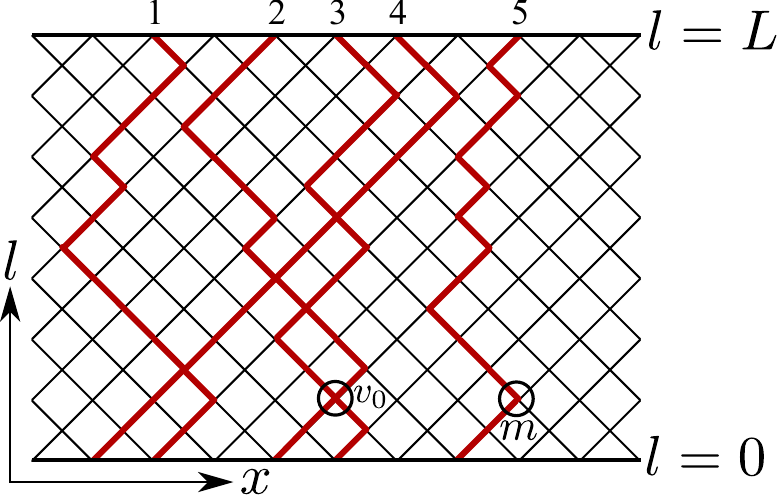}
	\caption{
\as{\textsl{(Color Online)}} Schematic representation of the model. Directed lines, which are indistinguishable, are labeled $1\ldots5$.
The weight associated with each line configuration depends on the number of turns, with each turn being assigned weight $m$.
Each interaction vertex is assigned a weight controlled by the interaction strength $v_0$.}
\label{fig:model}
\end{figure}

To identify the quantum analog appropriate to a single zigzagging directed line,
we recall---following Feynman and Hibbs~\cite{FeynmanHibbs} (and further developments in Refs.~\cite{Gaveau1984, Jacobson1984, Gaveau1989})---that the (discretized) Feynman path-integral representation of the quantum propagator ${\cal Z}(C_f, C_i)$ between the initial state $C_i$ and the final state $C_f$ of a relativistic particle 
consists of a sum over zigzagging paths.
For a particle of mass $\mu$ taking a path with $\tau$ turns, the amplitude is $(i \epsilon \mu)^\tau$ and the propagator is thus given by
\begin{equation}
{\cal Z}(C_f, C_i) = \lim_{\epsilon \rightarrow 0}\frac{1}{2 \epsilon}\sum (i \epsilon {\mu})^{\tau},
\label{eq:prop}
\end{equation}
where the sum is taken over all paths consistent with the initial and final states; $\epsilon$ is the lattice constant (i.e., the polymer Kuhn length); 
and we have chosen units such that $\hbar =  c = 1$.
The state $C$ refers to the combination of the positional coordinate $x$ of the particle and its motion direction $\sigma$ ($\equiv\pm$).
As we are interested in polymer phenomena on length scales much larger than the Kuhn length, we now take the continuum limit $\epsilon \rightarrow 0$. 
It has been shown~\cite{FeynmanHibbs} that, in this limit, for a particle propagating for time $L$,
${\cal Z}$ is given by the matrix element $\langle C_f | e^{- i L \hat{h}} | C_i \rangle$, where 
the $(1+1)$-dimensional Dirac Hamiltonian $h$ is given, in the Weyl representation, by $- i \sigma_z \partial_x - \mu \sigma_x$, in which
\begin{equation}
\sigma_x \equiv \left(\begin{array}{cc}
    0&1 \\
    1&0 \end{array}\right) 
\mbox{ and } \, \sigma_z \equiv \left(\begin{array}{cc}
    1&\phantom{-}0 \\
    0&-1 \end{array}\right)
\end{equation}
are the Pauli matrices, and the discrete degree of freedom is the direction of motion.

For a zigzagging directed line, the classical partition function $Z(C_f, C_i)$ associated with the line statistics has a form similar to ${\cal Z}(C_f, C_i)$, i.e.,
\begin{equation}
Z(C_f, C_i)=\frac{1}{2 \epsilon}\sum (\epsilon m)^{\tau}.
\end{equation}
\as{Here, we interpret the weight $\epsilon m$ as a Boltzmann factor, which has an exponential dependence on the reciprocal of the temperature.} 
Replacing $i \mu$ by $m$ in Eq.~(\ref{eq:prop}) transforms ${\cal Z}$ into Z. Therefore, in the continuum limit
we have that $Z = \langle C_f | e^{- L \hat{H}} | C_i \rangle$, in which the imaginary-time Dirac Hamiltonian $H$ in the Weyl representation is given by
\begin{equation}
\label{eq:H}
H = m I + \sigma_z \partial_x - m \sigma_x = \left(\begin{array}{ll}
     \partial_x + m & \, \phantom{- \partial_x} - m \\                                          
     \phantom{\partial_x} - m & \, - \partial_x + m
     \end{array}\right).
\end{equation}
Here, we have added the term $m I$ to $H$, where $I$ is the identity.
This term has the effect of shifting the eigenspectrum by a constant. \as{Although the statistics of each polymer is not affected by the $m I$ term, this term is necessary if we are to arrive at the continuity equation, examined below. In turn, this arrival is necessary for the interpretation of the imaginary-mass Dirac equation as an equation for the probability density of a single particle.}

By analogy with the well-known equivalence between the Feynman path integral and time-dependent Schr\"odinger equation,
we may analyze the statistics of the zigzagging directed line in terms of the time-dependent imaginary-mass Dirac equation 
\begin{equation}
\label{eq:imdirac}
- \partial_{l} \Psi = H \Psi
\end{equation}
 for the doublet $\Psi$.
We note that the Hamiltonian $H$ has been considered in Refs.~\cite{Kholodenko1990,Kholodenko1998} in connection with polymer solutions.
The orientation of a polymer segment (see Fig.~\ref{fig:model}), which we shall call its \emph{tilt}, corresponds to the velocity of the quantum particle.
It is associated with the operator $d \hat{x} / dl= [\hat{H}, \hat{x}] = \hat{\sigma}_z$.
The thermal average of the polymer direction is therefore captured by the quantum expectation value $\langle \hat{\sigma}_z \rangle$.

To pass to the nonrelativistic (i.e., large-mass) limit of the imaginary-mass Dirac equation, we note that it reduces to a single second-order partial differential equation---the telegrapher's equation---which, combined with the continuity equation, $\partial_l (\Psi_1+\Psi_2) + \partial_x (\Psi_1 - \Psi_2) = 0$, forms an equivalent description. Its solution is known~\cite{Masoliver1996}, and captures a combination of diffusive and propagating (i.e., wave-like) motion.
In the aforementioned limit, the telegrapher's equation reduces to the diffusion equation $\partial_l \Psi_{1,2} = (2m)^{-1} \partial_x^2 \Psi_{1,2}$~\cite{f1,Goldstein1951,Masoliver1996}, i.e., the quantum analog is governed by the (nonrelativistic) Schr\"{o}dinger Hamiltonian $(2m)^{-1} \partial_x^2$.
To summarize, in the real-time nonrelativistic limit the Dirac equation becomes the Schr\"odinger equation, whereas in the imaginary-time nonrelativistic limit the telegrapher's equation becomes the diffusion equation~\cite{Gaveau1989}.

To analyze the many-polymer liquid, we make use of results for the one-polymer system,
which follow from the separation-of-variables solution of the imaginary-mass Dirac equation.
Thus, we hypothesize the separated solutions
\begin{equation}
{\boldsymbol\Psi}_{nk}(x; l) = \re\, \left( {\boldsymbol\psi}_{nk} \, e^{- l E_{nk}} \, e^{i k x} \right),
\label{eq:sol}
\end{equation}
where $k$ is the (real) momentum, and the eigenvalues $E_{nk}$ and constant tilt-doublets ${\boldsymbol\psi}_{nk}$ may, in general, be complex.
By inserting Eq.~(\ref{eq:sol}) into the imaginary-mass Dirac equation~(\ref{eq:imdirac}), we arrive at the reduced eigenproblem:
\begin{equation}
\label{eq:H2}
\left(\begin{array}{ll}
     i k + m - E_{nk} & \phantom{-i k} - m \\
     \phantom{i k}- m & - i k + m - E_{nk}
     \end{array}\right)
{\boldsymbol\psi}_{nk} = 0,
\end{equation}
which has nontrivial solutions provided $E_{nk} = m \mp \sqrt{m^2 - k^2}$ (for $n = 1$, $2$). 
For $k < m$, $E_{nk}$ are real and the separated solutions have the form
\begin{equation}
{\boldsymbol\Psi}_{\!nk}(x; l) \! = \! \big( \re\, {\boldsymbol\psi_{nk}} \big) e^{-l E_{nk}} \! \cos k x  
\! - \! \big( \im\, {\boldsymbol\psi_{nk}} \big) e^{- l E_{nk}} \! \sin k x , \nonumber
\end{equation}
whereas for $k > m$ they have the distinct form
\begin{align}
{\boldsymbol\Psi}_{nk}(x; l) &= \big( \re\, {\boldsymbol\psi}_{nk} \big) e^{- l m} \cos \left(k x \mp l \sqrt{k^2 - m^2}\right)  \nonumber \\
 & \, - \big(\im\, {\boldsymbol\psi}_{nk} \big) e^{- l m} \sin \left( k x \mp l \sqrt{k^2 - m^2}\right), 
\end{align} 
resulting from $E_{nk}$ being complex.
The latter form results from the fact that the imaginary-mass Dirac Hamiltonian is non-Hermitian,
and thus has both real and imaginary parts to its eigenvalue spectrum.
Such a non-Hermitian Hamiltonian does not describe an isolated quantum system, although
analogous Hamiltonians can be used to model certain open quantum systems~\cite{Bender2002}.
Non-Hermitian Hamiltonians have also been applied to sheared directed line liquids under tension, via the mapping to a nonrelativistic quantum system subject to an imaginary vector potential~\cite{Hatano1996,Hatano1997,Hatano1998}.
When we come to study the response of zigzagging directed line liquids to shear stress,
we shall---following Refs.~\cite{Hatano1996,Hatano1997,Hatano1998}---introduce the stress $s$ coupled to the tilt operator $\sigma_z$, so that $H \rightarrow H + s \sigma_z$; 
equivalently, $k \rightarrow k - i s$, i.e., here, too, $s$ serves as an imaginary vector potential. 

We pause to mention a well-known simplification that holds for bulk properties of long systems,
viz., ground-state dominance. Using bra-ket notation, 
a polymer segment localized at position $x$ with tilt label $\sigma$ corresponds
to the quantum state $| x \sigma \rangle$. 
We then define the eigenstates $| n k \rangle = \int dx \sum_{\sigma} | x \sigma \rangle \, {\psi}_{nk}(\sigma) e^{i k x}$, and express the transfer-matrix operator appearing in the partition function $Z$
in the eigenstate basis as follows:
\begin{equation}
e^{-L \hat{H}} = \sum_{n,k} | n k \rangle \, e^{- L E_{nk}} \, \langle n k | \approx | \textrm{gs} \rangle \, e^{- L E_{\textrm{gs}}} \, \langle \textrm{gs} |,
\label{eq:gsdom}
\end{equation} 
where $| \textrm{gs} \rangle$ denotes the eigenket for which $\re(E)$ is minimized.
The series in Eq.~(\ref{eq:gsdom}) is indeed dominated (for large $L$) by a single ground state, provided the ground state is nondegenerate. 
By using the expression $F = - T \ln Z$ for the thermodynamic free energy $F$ of the polymer system, together with Eq.~(\ref{eq:gsdom}), we obtain $F = T L E_{\textrm{gs}}$.
For a single zigzagging directed line, the ground state
is $(n,k) = (1,0)$, so that $E_{\textrm{gs}} = 0$, \as{and the components of the tilt-doublet are
$\left({\psi}{(+)},{\psi{(-)}}\right) = (1,1) /\sqrt{2}$}.

Prior to giving our analysis of the many-polymer liquid, we note a simple, self-consistent variational argument for the many-body ground state energy $E_{\textrm{gs}}$ of the interacting liquid
consisting of $N$ polymers at mean density $\rho$.
As a consequence of thermal fluctuations, each polymer occupies a characteristic width $w$ in the $x$-direction. 
In analogy with work on lamellar smectics~\cite{Golubovic1989,ChaikinLubensky}, we replace the interacting many-line system by a single line in a harmonic external potential. 
For the relativistic system, the Dirac oscillator [which is defined using a linear potential term, $V(x) = x \sigma_z$] is a solvable analog of the nonrelativistic harmonic oscillator~\cite{Moshinsky1989, Villalba1994,Szmytkowski2001}. Replacing the mass of the Dirac oscillator by an imaginary value, its energy eigenvalues 
acquire the form $E_{\lambda\mp} = m \mp \sqrt{m^2 - 2 |\lambda_{\mp}| w^{-2}}$ (for positive integers $\lambda_-$ and nonnegative integers $\lambda_+$)~\cite{Villalba1994,Szmytkowski2001}.
Thus, given the lowest-energy state $\lambda_- = 1$ and the self-consistency condition $\rho = w^{-1}$, we conclude that
\begin{equation}
\label{eq:self-cons}
E_{gs}/N = E_{1-} = m - \sqrt{m^2 - 2 \rho^2},
\end{equation}
from which the polymer liquid free energy follows.
Note that this self-consistent spectrum breaks down as $\rho \rightarrow \sqrt{2} m/2$, i.e., the value at which the square root in Eq.~(\ref{eq:self-cons}) becomes imaginary.

Returning to our main goal, we now formulate the statistical mechanics of the zigzagging directed polymer liquid in terms of quantum many-particle physics. To do this, we choose the polymers to be identical (i.e., indistinguishable).
We focus on joint probability densities for the polymer segments that, in the quantum analogy, correspond to the real and positive subset of wavefunctions $\Psi(X;\Sigma)$ [where $X \equiv (x_1,\ldots,x_N)$, and $\Sigma \equiv (\sigma_1,\ldots,\sigma_N)$].
Specifically, we consider a polymer liquid in which two segments that are tilted in the same direction
cannot occupy the same location in space. Thus, two polymers cannot ``comove\rlap,'' but they may cross~(see Fig.~\ref{fig:model}). For each crossing, we assign a statistical weight controlled by an interaction strength $v_0$. In a quasi-two-dimensional polymer system, the polymers may be able to pass over and under each other, with a corresponding energy cost. However, this energy may be sufficiently large such that two polymer staying on top of each other (i.e., comoving) is highly improbable.
In the quantum analogy, this system corresponds to the sector of many-body quantum states that are symmetric under pairwise exchange of particles: $|x_1,x_2,\ldots;\sigma_1,\sigma_2,\ldots\rangle = |x_2,x_1,\ldots;\sigma_2,\sigma_1,\ldots\rangle$ etc., i.e., 
bosonic states, with the additional constraint that $|x_1,x_1,\ldots;\sigma_1, \sigma_1,\ldots\rangle = 0$ etc.

Within the quantum analogy, this system can be usefully addressed using Girardeau's extension~\cite{Girardeau2004} of his 1D mapping~\cite{Girardeau1960}
between hard-core bosons and noninteracting fermions.
For a system possessing only \emph{spatial} degrees of freedom~\cite{DeGennes1968,Rocklin2012,Rocklin2013} (i.e., without tilt), this mapping proceeds by using the factor $B(X) = \prod_{n<m}\sgn(x_n - x_m)$ in the transformation $|X\rangle_B = B(X)|X\rangle_F$ between symmetric (boson, B) and antisymmetric (fermion, F) states.
The Pauli principle, operative for the Fermi states, translates into a local hard-core repulsion for the Bose states, so that two bosons are excluded from occupying the same point in space~\cite{Girardeau1960}.
This mapping may be extended to states with tilt, using 
$| X,\Sigma\rangle_B = B(X)|X,\Sigma\rangle_F$,
i.e., by introducing a sign of $-1$ into the fermionic state under the exchange of the position (but not tilt) labels~\cite{Girardeau2004}.
Thus, the bosonic state $|X,\Sigma\rangle_B$ indeed vanishes if $(x_i,\sigma_i) = (x_j, \sigma_j)$ for any pair of particles; but it need not vanish if $\sigma_i \ne \sigma_j$.

To explore structure in the polymer liquid, we now derive the corresponding quantum many-particle Hamiltonian in the occupation number representation.
First, we define the states $|\!\{\alpha_{nk}\}\!\rangle \! =\! \prod_{nk} \! \big( \hat{c}^\dagger_{nk} \big)^{\alpha_{nk}} \, |0\rangle$, 
where $\alpha_{nk}$ are the particle occupation number of the states $|nk\rangle$, 
$\hat{c}^\dagger_{nk}$ are the corresponding creation operators, 
and $|0\rangle$ is the no-particle state. 
As we are considering Fermi statistics, the operators $\hat{c}^\dagger_{nk}$ obey anti-commutation relations. 
Using the eigenstates $\psi_{nk}(\sigma)$ of the single-particle Hamiltonian, we construct the field operators ${\hat{\psi}_\sigma}(x) = \sum_{nk} 
\psi_{nk}(\sigma) \, e^{i k x}\, \hat{c}_{nk}$, in terms of which the many-body Hamiltonian $\hat{\cal H}$ has the form (c.f.~Eq.~\ref{eq:H})
\begin{equation}
\label{eq:Hmb}
\hat{\cal H} = \int dx \left\{ \sum_{\sigma\sigma^\prime}\hat{\psi}_{\sigma}^\dagger \, H_{\sigma\sigma^\prime}^{\phantom{\dagger}}
 \, {\hat{\psi}_{\sigma^\prime}}^{\phantom{\dagger}}
+ v_0 \, {\hat{\psi}_{+}}^\dagger \, {\hat{\psi}_{-}}^\dagger \, \hat{\psi}_{-}^{\phantom{\dagger}} \, \hat{\psi}_{+}^{\phantom{\dagger}} \right\},
\end{equation}
where all of the field operators have argument $x$.
Next, by analogy with quantum statistical mechanics, we introduce $l$-dependent operators
$\hat{\psi}_\sigma(x; l) \equiv e^{\hat{\cal H} l} \, \hat{\psi}_\sigma(x)  \, e^{- \hat{\cal H} l}$
and the partition function $Z \equiv \mathrm{Tr} \, e^{-L \hat{\cal H}}$ (where the trace $\mathrm{Tr}$ is taken over all $N$-particle states), and thermal expectation values
of operators $\hat{O}$ are given by $Z^{-1} \mathrm{Tr} \, \hat{O} \, e^{- L \hat{\cal H}}$.

\begin{figure}
     \includegraphics{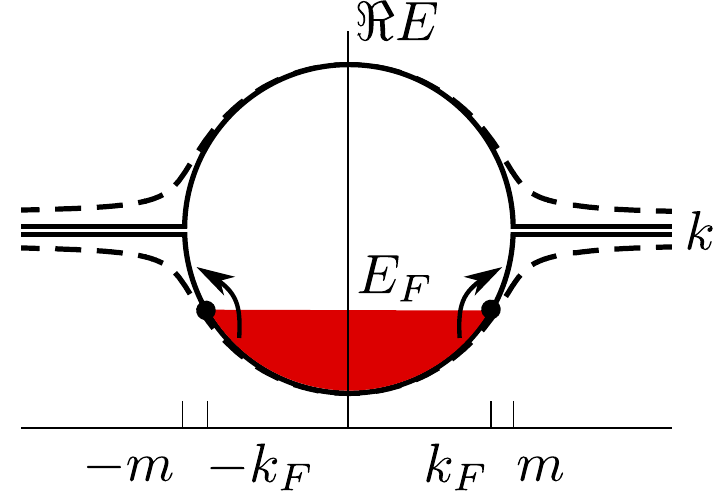}
	\caption{\as{\textsl{(Color Online)}} Single-particle energy spectrum of the fermion analogy in the absence (solid lines) and presence (dashed lines) of an external shear stress. Note the large degeneracy for $|k| \ge m$, which the shear eliminates.
In the $v_0 = 0$ case, the many-body ground state
is constructed by filling the Fermi sea (red).
For $\rho < m/\pi$, the low-energy excitations are of the particle-hole type (shown by arrows), and involve single-particle states near the Fermi points $k = \pm k_F$.}
\label{fig:bands}
\end{figure}

Thus, the zigzagging polymer liquid maps to an interacting $N$-fermion fluid described by the Hamiltonian~(\ref{eq:Hmb}). 
For the case $v_0 = 0$, the exact ground state of this Fermi fluid has only the $N$ lowest-energy single-particle states occupied.
The nature of this ground state changes qualitatively as the value of the Fermi wavevector $k_F$ (which is related to the density via $\rho = k_F/\pi$) is increased to $m$.
For $k_F < m$, the Fermi sea fills the bottom portion of the spectrum and the real part of the total energy is given by $\re E = m N -  \sum_{k<k_F} \sqrt{m^2 - k^2}$~(see Fig.~\ref{fig:bands}).
For $k_F > m$, there are many degenerate ground states, all with $\re E = m N - \sum_{k<m}\sqrt{m^2 - k^2}$; in these states all single-particle states with $n = 1$ and $k < m$ are occupied, with the rest of the occupied states being arbitrarily chosen from the doubly-degenerate band that has $\re E = m$ and $k > m$~(see Fig.~\ref{fig:bands}).
For $k_F > m$, a unique ground state can be achieved by imposing a shear stress $s$~(see Fig.~\ref{fig:bands}).
Thus, for $s \ne 0$, the ground state is non-degenerate for any particle density.

\begin{figure}
     \includegraphics{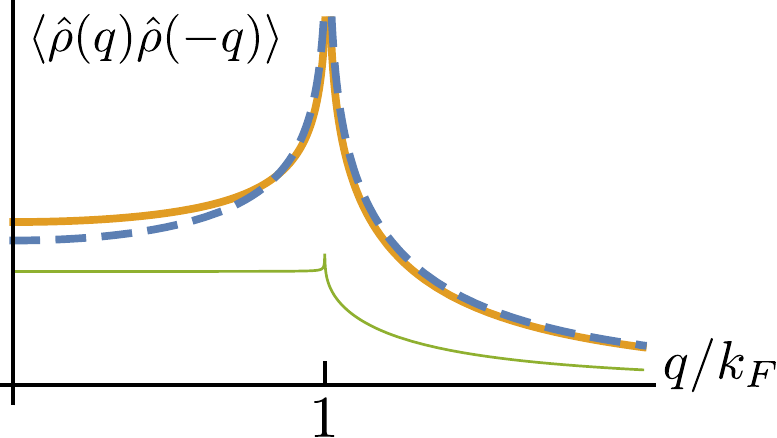}
	\caption{\as{\textsl{(Color Online)}} Wavevector-dependent zigzagging polymer density correlation function~$\langle \hat{\rho}(q) \hat{\rho}(-q) \rangle $ for $\omega = 0$ (i.e., wavevector transverse to the preferred polymer direction) with $v_0 = 0$. The cases $k_F/m = (0, 0.5, 0.99)$ are shown using, respectively, dashed, thick, and thin lines.
Whereas in the low-density limit, the result for directed polymers without bending rigidity, i.e., from Ref.~\cite{DeGennes1968}, is recovered, at higher densities, the polymer bending rigidity has an effect on the structure of the polymer liquid. Furthermore, at $k_F = m$, the Kohn-type anomaly at $q = k_F$ changes its character.}
\label{fig:corr}
\end{figure}

Using these fermion ground states, we calculate the correlations of density in the zigzagging directed line liquid with $v_0 = 0$ using the exact expression
\begin{equation}
 \langle \hat{O}(q,\omega) \hat{O}(-q,-\omega) \rangle \!= \! \re \sum_{knn^\prime} \frac{O_{n(q+k)} - O_{n^\prime k}}{E_{n(q+k)} - E_{n^\prime k} - \omega},
\nonumber 
\end{equation}
where $\hat{O}$ is an operator (e.g., $\hat{\rho} \equiv \sum_{\sigma}\hat{\psi}_\sigma^\dagger \hat{\psi}^{\phantom{\dagger}}_\sigma$), 
$O_{nk}$ is the expectation value of $\hat{O}$ in the single-particle
state $nk$. 
We compute these correlation functions using the complex basis~(\ref{eq:sol}).
For $\omega = 0$, the results are plotted in Fig.~\ref{fig:corr}.

\begin{figure}
     \includegraphics{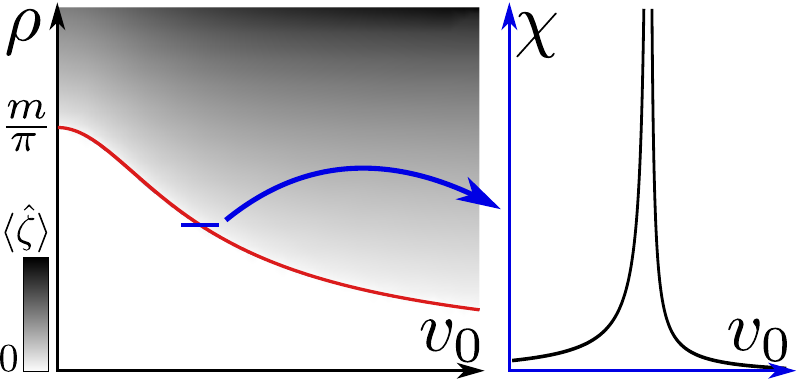}
	\caption{\as{\textsl{(Color Online)}} Left: Phase diagram in the plane spanned by density $\rho$ and interaction $v_0$, obtained using the self-consistent solutions of the tilt in the ground state.
The density plot shows the degree of tilt order $\langle \hat{\zeta} \rangle$ in the titled phase, i.e., above the phase boundary line. Right: Uniform tilt susceptibility near the phase boundary, exhibiting a divergence characteristic of a continuous phase transition.}
\label{fig:sc}
\end{figure}

Next, we study the uniform tilt susceptibility in analogy with the Stoner theory of itinerant magnetism, which addresses a repulsive spin-\nicefrac{1}{2} Fermi gas~\cite{Stoner1938}.
To do so, we use a self-consistent approximation and ignore fluctuations, which we recognize destroy true long-range order in one-dimensional Fermi systems---instead, we expect the true ground state to have long-range tilt correlations.
For zigzagging directed lines, the self-consistent equation (SCE) results from the Hartree-Fock approximation of the interaction:
$v_0 \langle  \hat{\psi}_-^\dagger\hat{\psi}_- \rangle \hat{\psi}_+^\dagger\hat{\psi}_+ + v_0 \langle \hat{\psi}_+^\dagger \hat{\psi}_+ \rangle \hat{\psi}_-^\dagger\hat{\psi}_-$~\cite{FetterWalecka}.
Rewriting this expression in terms of the expectation value $\langle \hat{\zeta} \rangle$ of the tilt operator $ \hat{\zeta} \equiv \hat{\psi}_+^\dagger\hat{\psi}_+ - \hat{\psi}_-^\dagger\hat{\psi}_-$,
the relevant term becomes $- s_{\mathrm sc} \hat{\zeta}$, where
\begin{equation}
\label{eq:sce}
s_{\mathrm sc} = \frac{v_0}{2} \langle \hat{\zeta} \rangle,
\end{equation}
which is the SCE\as{, expressed in terms of self-consistent field $s_{\mathrm sc}$}. We evaluate $\langle \hat{\zeta} \rangle$
in the ground state by using a (nonorthogonal) transformation to the eigenbasis of the Hamiltonian~\cite{f2} to obtain
\begin{equation}
\langle \hat{\zeta} \rangle = \frac{1}{\pi} \im \left( \sqrt{m^2 - (k_F + i s_{\mathrm sc})^2} \, \right).
\end{equation}
Substituting this expression into the SCE~(\ref{eq:sce}),
we find a non-trivial solution if and only if $(1 + g^2 )k_F^2 > m^2$, where $g \equiv v_0/2 \pi$~(see Fig.~\ref{fig:sc}).
In this region, the SCE (with its neglect of fluctuations) predicts a tilt of the polymer system in equilibrium~(see Fig.~\ref{fig:sc}).
Note that for $k_F > m$, this solution exists even for the case $v_0 = 0$; this results from the fact that an applied stress $s$ breaks the degeneracy between the two bands, as shown in Fig.~\ref{fig:bands}, leading to polymer-system tilt that increases with density beyond $m/\pi$.
The critical line shown in Fig.~\ref{fig:sc} corresponds to an Ising-like continuous phase transition in the mean-field approximation. Accordingly: {(i)}~the Ising order parameter (i.e., the tilt) scales as the square root of the distance to the critical line, $\langle \hat{\zeta} \rangle = \pi^{-1}\sqrt{k_F^2 - m^2(1 + g^2)^{-1}}$; {(ii)}~on the critical line, the tilt scales with the shear stress as 
$\langle \hat{\zeta} \rangle \sim \frac{1}{\pi} \left( \frac{2 g m^2}{1 + g^2}\right)^{1/3} s^{1/3}$; and {(iii)}~the uniform tilt susceptibility diverges on approach to the critical line~(see Fig.~\ref{fig:sc}b), as seen in the expression
\begin{equation}
\chi  = \left\{
     \begin{array}{cl}
       {\displaystyle \frac{1}{\pi}\frac{k_F}{\sqrt{m^2-k_F^2}-g k_F}} & \mbox{for } 
\langle \hat{\zeta} \rangle = 0; \\
\noalign{\smallskip}
       {\displaystyle \frac{1}{\pi}\frac{m^2 g}{1+g^2}\frac{1}{(1+g^2)k_F^2-m^2}} 
& \mbox{for } 
 \langle \hat{\zeta} \rangle \ne 0.
     \end{array}
   \right.
\end{equation}

The thermodynamics of the zigzagging directed line liquid follow from the polymer free energy $T L E_{gs}$.
For example, for $k_F < m$, $g = 0$, and $s = 0$, the compressibility is given by $T \sqrt{m^2 - (\pi \rho)^2}/ (\pi^2 \rho^3)$.
Its vanishing as $\rho \rightarrow m / \pi$ is an additional signature of the cross-over to a tilted state.

We note that the model described here addresses the structure of the zigzagging polymer liquid on mesoscopic length scales, 
i.e., between the molecular structure of the polymer fluid and its larger-scale collective properties.
We also note that Eq.~(\ref{eq:Hmb}) is a variant of the Thirring model~\cite{Thirring1958}. 
As is well known, the original Thirring model can be solved exactly by using a mapping onto the
bosonic sine-Gordon model~\cite{Coleman1975}. However, using the same bosonization methods, the Hamiltonian~(\ref{eq:Hmb}) does not map onto a realizable Hamiltonian for a Bose system because the one-particle energy, Eq.~(\ref{eq:H}), maps onto a complex term. 
Thus, it remains an open question whether the many-body
problem analyzed here is exactly solvable.

Let us also comment on the role of dimensionality in this model.
(i)~It would be interesting to explore whether extensions of the
Hamiltonian~(\ref{eq:H}) could be applied to higher-dimensional systems of directed lines. 
(ii)~Although possible~\cite{Souslov2013}, the transmutation of quantum statistics in higher-dimensional systems requires a more elaborate construction.  
(iii)~In two-dimensional directed polymer liquids, fluctuations eliminate true long-range order, though vestiges of it remain in the form of long-range correlations.

We conclude with a discussion of the physical content of the Dirac polymer liquid model. This model indicates the possibility of a new type of order (i.e., \emph{tilt} order) in a liquid of directed polymers having bending rigidity. Within such a liquid, we find that this tilt order results from an interplay between bending rigidity and strong polymer-polymer repulsion. To treat this repulsion, we use an analogy between directed polymers and quantum particles. Within this analogy, the orientational degree of freedom of a polymer segment maps onto the intrinsic spin of a quantum particle. The analogy enables us to examine the directed polymer liquid using techniques known from the study of one-dimensional, two-band Fermi liquids. As a result, we are able to describe the emergence of order in the polymer liquid in terms of a Stoner-like theory of itinerant magnetism~\cite{Stoner1938}, subject to fluctuations. It is noteworthy that, for sufficiently high polymer-densities, even weak interactions lead to tilt order. Thus, for the polymer liquid, we find phenomenology quite reminiscent of that central to Onsager's description~\cite{Onsager1949} of the isotropic-nematic transition. We emphasize the necessity of sufficient polymer rigidity for the emergence of tilt order that we have discussed here.

\emph{Acknowledgments -- } We thank Randall Kamien, Tom Lubensky, and Rafael Hipolito for fruitful discussions about this work.
This work was supported by NSF DMR 12 07026 and by the Georgia Institute of Technology. One of us (BL) would like to thank CONICYT and Becas Chile for their financial support.


\begin{thebibliography}{35}
\bibitem{f1}
More rigorously, this requires taking the limit $c \rightarrow \infty$ in such a way that $c^2/2 m \equiv D$ is a finite diffusion constant.
\bibitem{f2}
We derive identical expressions either by considering the ``group velocity'' of the ground state $\im d E_{gs}/d k_F$ or by substituting the SCE into Eq.~(\ref{eq:Hmb}) to obtain an expression for the mean-field free energy
of zigzagging directed lines, which we then minimize w.r.t. $s_{\mathrm sc}$.
\end{thebibliography}
\end{document}